%% file: aic_template.tex
\documentclass[twocolumn]{article}

\usepackage{amsmath}
\usepackage{comment}
\usepackage{float} 
\usepackage[ruled,vlined,norelsize,\languagename]{algorithm2e}
\usepackage{color}						

\usepackage{ctable}
\usepackage{spreadtab}
\usepackage{longtable}
\usepackage{multirow}
\usepackage{subcaption}

\begin{document}
\title{Norms for Beneficial A.I.: A Computational Analysis of the Societal Value Alignment Problem}


\author{{Pedro M.} {Fernandes}, {Francisco C.} {Santos},{Manuel} {Lopes}\\
{pedro.miguel.rocha.fernandes,franciscocsantos,manuel.lopes}@tecnico.ulisboa.pt\\
{INESC-ID and Instituto Superior Técnico, {Univ. de Lisboa}, Lisbon, {Portugal}}}

\maketitle

\begin{abstract}
The rise of artificial intelligence (A.I.) based systems is already offering substantial benefits to the society as a whole. However, these systems may also enclose potential conflicts and unintended consequences. Notably, people will tend to adopt an A.I. system if it confers them an advantage, at which point non-adopters might push for a strong regulation if that advantage for adopters is at a cost for them. Here we propose an agent-based game-theoretical model for these conflicts, where agents may decide to resort to A. I. to use and acquire additional information on the payoffs of a stochastic game, striving to bring insights from simulation to what has been, hitherto, a mostly philosophical discussion. We frame our results under the current discussion on ethical A.I. and the conflict between individual and societal gains: the societal value alignment problem. We test the arising equilibria in the adoption of A.I. technology under different norms followed by artificial agents, their ensuing benefits, and the emergent levels of wealth inequality. We show that without any regulation, purely selfish A.I. systems will have the strongest advantage, even when a utilitarian A.I. provides significant benefits for the individual and the society. Nevertheless, we show that it is possible to develop A.I. systems following human conscious policies that, when introduced in society, lead to an equilibrium where the gains for the adopters are not at a cost for non-adopters, thus increasing the overall wealth of the population and lowering inequality. However, as shown, a self-organised adoption of such policies would require external regulation.

\end{abstract}





\input{Main_Text.tex}

\section*{Acknowledgements}
This research was supported by FCT-Portugal through grants UID/CEC/ 50021/2019, PTDC/EEI-SII/5081/2014, PTDC/MAT/STA/3358/2014, and by the EU H2020 RIA project iV4xr : 856716.





\bibliographystyle{plain}           
\bibliography{Bibliography}  

%

\end{document}

%% file: Main_Text.tex

\section{Introduction}

Several applications already have an Artificial Intelligent system (A.I.) taking decisions in place of their owners. It is expected that in the future, such delegation of decisions will become more ubiquitous and effective. It is still open to debate whether that will have a positive or a negative impact on society \cite{yudkowsky2008artificial, brundage20184}. Strong voices highlight the dangers of A.I. \cite{armstrong2016racing} and call for regulation \cite{guardian18cambridAIregulNEWS}, some others dismiss such fears \cite{technologyreview17brooksAINEWS} and are against regulation \cite{forbes17airegulationNEWS}.  Some of these discussions come from a lack of understanding of the current A.I. capabilities and strong divergences about its future developments, especially in artificial general intelligence (AGI). Some concerns might be true when AGI is created but not under the current state-of-the-art. How fast we can get there is still open to debate \cite{bostrom2014superintelligence}, and so is if we should strive to get there fast, or delay it \cite{bostrom2003ethical, good1966speculations}. But even under the current state-of-the-art in A.I. there are problems that may arise with their introduction, e.g. autonomous vehicles \cite{bonnefon2016social}, automatic hiring systems \cite{sanchez2020does} and stock exchange \cite{mcnamara2016law}.

A strong regulation could decide that A.I. systems should act using an egalitarian or utilitarian perspective. However, a utilitarian perspective or norm might not be efficient, and in many cases, an egalitarian solution does not exist. An utilitarian A.I., which would strive to maximize the overall utility gain of the world, would often have to act against the interests of its owner. If people can choose to adopt or not an A.I. system, we can expect they will only do it if it is \textit{individually rational} to do so. In principle, either the A.I. system gives an individual advantage for its owner, or it will not be bought. If there is no interest in buying, there will be no interest in production, curbing research and development.

Without any regulation, A.I. systems might lead to invasion of privacy, use of confidential information, cheating in games, collusion in public contracts, and many others. Even if legal their efficiency and effectiveness might greatly unbalance the societal scales, increasing the inequality in wealth distribution. In this case, we can expect that non-adopters might push for a substantial regulation of A.I. or even for its abolition. 

It is thus challenging to conciliate these two goals of aligning the preferences of A.I. adopters and those of the non-adopters. We call this the \textbf{societal value alignment} problem. Besides being advantageous for the adopters of A.I. (\textit{individual rationality}), it needs to be better for the non-adopters and so for everyone in the society (\textit{societal rationality}). The rise of A.I. systems has the power to create novel market dynamics \cite{tan2018ai} and challenges. Efforts should be made to model these possible future worlds, so that we understand them better before we are in the midst of the problems that might arise. Voices in the scientific community begin to pressure for the research on this area, and on the ethical, scientific and engineering problems it presents \cite{taylor2016alignment, russell2015research, Han2019AIWar, pereira2016programming}. 
~\\

Here we aim to model how A.I. systems can provide an advantage for those adopting them (fostering \textbf{adoption} and creating incentives for the scientific, technological and societal development) but without creating such advantage at the expense of others (allowing for societal \textbf{acceptance} of the systems). \textbf{Adoption} is here defined as the delegation of decisions to an A.I. system that is done by an individual. \textbf{Acceptance} concerns the societal opinion regarding A.I. systems, which will be highly influential on their legislation. To do so, we define several different types of A.I. systems, adopting different types of norms, ranging from pure selfish to pure utilitarian. Then we study the time evolution of the adoption of each type of A.I. when they compete against each other and also the equilibrium for each A.I. system in particular.

Individual adopters of A.I. systems can be seen as: singular citizens adopting A.I. systems for personal gains (e.g. a person buying an autonomous vehicle); corporations buying A.I. to increase profits (e.g. a company adopting an A.I. based hiring system); political entities using A.I. to gain influence (e.g. political parties using A.I. algorithms in social media to influence elections); or even countries, deploying A.I. to gain an upper hand on war and trading. Our model abstracts individuals as equally complex entities that interact between each other, gaining or losing utility on each interaction. Utility is here used as an abstraction of something desirable/useful and could be seen, for example, as monetary currency, the strengthening of political positions or improvement of individual well-being. 

In particular, this analysis aims at answering the following research questions:
\begin{enumerate}
\item Will self-regarding individuals adopt A.I. systems?
\item With different types of A.I. systems available, which ones will be adopted?
\item If adopted, what is the individual and collective gain, depending on the strategy adopted by the A.I. system? 
\end{enumerate}

Based on the answers to these questions we then discuss the kind of regulation that might be needed to improve the individual and societal rationality of A.I. systems. These allow us to provide novel insights on the following questions:

\begin{enumerate}
\item Is any type of A.I. both acceptable and adoptable?
\item Taking into account all the evaluation criteria, including individual and societal, is it possible to create mechanisms/properties that improve all of them?
\item Considering even the extreme case where everyone uses the same A.I. enabled system, will they obtain the same benefit?
\end{enumerate}

This paper is organized as follows. In Sec. 2 we discuss the related work. Sec. 3 introduces the main contribution of our work and details the set of possible behaviours or norms that a A.I. system may have, and how we evaluate their performance when acting in a population comprising humans and artificial systems. To do so, we introduce a novel game theoretic model for the dynamics of adoption of A.I. systems. Section 4 presents the results of our computer simulations. In Sec. 5 we introduce another simulation aiming at understanding the inequality emerging in the cases where everyone has the same A.I. system. In Sec. 6 we summarize our conclusions and, finally, in Sec. 7, we present a more extended discussion about the impact of this study in the dilemmas related to the introduction of A.I. systems.

\section{Related Work}
\label{sec:related}

The problem of aligning one A.I. system with all the individuals of a society (A.I. and non A.I.), knowing many of these individuals might have contradictory values, is a complex one. Even if the system is capable of perfectly aligning with each individual, there is no perfect solution as in many situations, it won't be possible to be aligned with all of the individuals simultaneously. The system will have to choose with whom to be and not be aligned.

Many authors expect many negative effects from the adoption of A.I. systems and so many different ethical codes of conduct have been proposed. A code of conduct would represent the universal human values and by aligning with it, the A.I. system would be indirectly aligned with all humankind. First literary approaches include the famous laws of robotics by Isaac Asimov \cite{asimov2004robot}. They were natural language laws, which leads to obvious implementation problems, and in his books, Isaac Asimov proved his own laws flawed. After that, several ethical frameworks have been proposed. Some principles are found in almost all of them, others are characteristic of each approach. 

The Asilomar AI Principles \cite{asilomar2018principles} are a set of 23 principles intended to promote the safe and beneficial development of artificial intelligence. They have been endorsed by AI research leaders at Google DeepMind, GoogleBrain, Facebook, Apple, and OpenAI. Signatories of the principles include Elon Musk, the late Stephen Hawking, Stuart Russell, and more than 3,800 other AI researchers and experts. On August 30 of 2018, the State of California unanimously adopted legislation in support of the Asilomar AI Principles \cite{asilomarCalifornia}, taking a historic step towards A.I. research and development legislation.

The European Commission released a document by the high-level expert group on A.I., containing a set of ethic guidelines for trustworthy A.I. \cite{hleg2019ethics}. The document states that trustworthy A.I. should be lawful, ethical and robust. It further defines 7 key requirements that AI systems should meet in order to be deemed trustworthy. In the end, it presents a series of questions that entities should ask themselves to ensure they are meeting all the defined requirements. The Organisation for Economic Co-operation and Development (OECD) has released a legal instrument with recommendations on A.I., which was adopted by 42 countries \cite{oecd2019ethics}.

Floridi et al made a comparison of several ethical frameworks \cite{floridi2018an}, in which they analyse principles proposed by 6 different entities, including the previously mentioned Asilomar AI principles \cite{asilomar2018principles}, wielding 47 principles on total, and compare them to the existing 4 principles of bio-ethics (Non-maleficence; Justice; Beneficence; Autonomy) \cite{beauchamp2008principes}, finding a considerable overlap.
They argue that for the bio-ethics principles to be applied to the field of A.I., a fifth principle is needed: explicability. This principle incorporates both intelligibility and accountability. They go on to propose 20 action points, that is, recommendations for enabling a beneficial A.I. society.

Anderson et al defend that it may be possible to incorporate an explicit ethical component into a machine relying on inductive logic programming approach \cite{anderson2007machine}. The goal is to solve ethical dilemmas by finding ethical principles that best fit given positive and negative examples. They advocate the use of a modified version of the Turing test \cite{turing2009computing}, the comparative moral Turing test \cite{allen2000prolegomena}. This test is an elegant solution to the question ''What is an ethical/moral A.I. system?''. The test consists in giving to a human judge pairs of descriptions of actual, morally-significant actions of a human and an A.I. system. If the judge identifies the A.I. as a moral equal or superior to the human, then the A.I. system passed the comparative moral Turing test.

Conitzer et al describe a game theoretical approach to the problem of moral decision making \cite{conitzer2017moral}. Turning moral dilemmas into game-theoretic representation schemes, they then try to find the principles that guide human moral decision making. They also argue that such representations could be used alongside machine learning, which could lead to more reliable results and to the improvement of the representations.

However, some argue that having an ethical framework or even A.I. systems that pass the comparative moral Turing test is not enough \cite{yampolskiy2013artificial}. Roman Yampolskiy defends that it is insufficient to have a human-like morality on A.I. systems with super-human intelligence. On such systems, small moral mistakes, common in humans, could lead to the extinction of humanity. Furthermore, a moral A.I. system with super-human intelligence will be able to recursively self-improve, with no provided guarantees that the resulting improvements remain moral. Instead of an ethical approach, Yampolskiy proposes a safety engineering approach, able to provide proofs that developed A.I. systems will remain safe, even under recursive self-improvement \cite{yampolskiy2013safety}. Yampolskiy also proposes A.I. confinement as a possible approach while no safety guarantees are in place \cite{yampolskiy2012leakproofing, babcock2017guidelines}. This approach would consist in ensuring that an A.I. system could help humanity while having no ability to negatively influence the world around it. This idea of A.I. confinement had been first presented in \cite{drexler1986engines}, and discussed by Bostrom \cite{bostromoracle} and Chalmers \cite{chalmers2010singularity}. This is, however, more of a preventive measure than a perfect solution, as limiting the negative A.I. influence will also limit the possible positive influence.

The focus of most previous works was on considering high-level ethical principles for A.I. systems acting in a society or finding moral frameworks. In most cases there was no claim or prediction about the potential adoption of A.I. systems or their acceptance by non-adopters and by society in general. Just a few works considered the development of computational models on the impact of A.I.. For instance, one study analyzed the amount of safety precautions companies would take considering that they are competing with others for dominating A.I. \cite{armstrong2016racing}.

Taking a different stance from the majority of related literature, our work aims at understanding the dynamics of adoption (who chooses to use an A.I. system) and of acceptance (if non-adopters accept the use of A.I. by others) relying on computational models of population dynamics. A better understanding of such dynamics can both allow us to better predict the outcome of A.I. proliferation as well as inform future legislation to ensure a beneficial impact of A.I. technology in society.
Even though it is not yet a solved problem \cite{shapiro2002valuealignment}, for this paper, we will assume that an A.I. system can accurately estimate the goals of each individual with whom it interacts. With this assumption, we are able to study the problems that emerge at the societal level even after having the problem of individual value alignment solved.

\section{Methods} 
\label{sec:methods}

In this section, we present a game-theoretical framework to study the impact of the adoption of A.I. systems on individuals and on the society. We consider several different types of A.I. that might adopted, or allow to be used, by society. Some of them are purely social, others purely selfish. Although no exhaustive list is possible, we cover a set of different strategies to be able to study them in hybrid populations of A.I. and humans. We start by providing the model of a single interaction between two individuals, then explain how we model the difference in decision making between an A.I. system and a human, and finally present the simulated world and the used algorithms.

Henceforward, individual non-adopters of an A.I. system will be referred to as \textbf{H}, while individuals adopters of an A.I. system representative will be referred to as \textbf{A.I.}.

\subsection{Model of Interaction Between Individuals:}
\label{sec:interaction}

On each interaction between individuals, $I_1$ and $I_2$, a stochastic payoff matrix $M^t$ is generated. This is a $m$-by-$m$ matrix of payoff pairs that is different for every interaction. Being $a_1$ the action chosen by $I_1$ and $a_2$ the action chosen by $I_2$, the payoff received by each individual is respectively $u_1$ and $u_2$, such that:
\begin{eqnarray}
(u_1, u_2) = M^t(a_1,a_2).
\end{eqnarray}

In order to explicitly generate general sum games, as conclusions might be different in positive, negative or zero-sum worlds, the payoff matrices have the following structure:
\begin{eqnarray}
u_1 &=& R + z(0,2) |R| (\alpha - 1)\\
u_2 &=& -R + z(0,2) |R| (\alpha - 1)
\end{eqnarray}

Having $R = z(-3,3)$, where $z(a,b)$ represents a sample from a uniform distribution in the interval $[a,b]$. The interval $[-3,3]$ was chosen for the simulations, but any equivalent interval could be used. $R$ is the same for each $u_1$ and $u_2$ pair. $z(0,2)$ is applied independently for each element of the matrix. This $z(0,2)$ parameter creates an additional source of variability between different interactions, so that not all action pairs have the same overall utility gain. $|R|$ is the absolute value of $R$. We will call $\alpha$ an inflation constant. For $\alpha = 1$, the matrices will, on average, create a zero sum game where no payoff is created or lost, just transferred between individuals. For $\alpha > 1$ there is on average a positive total payoff, creating a positive sum game, and for $\alpha < 1$ there is on average a negative total payoff, creating a negative sum game. In our simulations, in order to study a positive sum world, we consider $\alpha = 1.2$. The number of possible actions per individual was set to $4$ $(m = 4)$, an empirically found balance between complexity and computational feasibility.

\subsection{Simulating A.I. Systems and Humans}
\label{sec:HVSR}

We now discuss how we can model an human acting versus an A.I. system. A.I. systems are different from humans and can provide several advantages: 
\begin{enumerate}
\item{\it A.I. systems are not be prone to fatigue or distractions and so will make less errors than humans:} As we model interactions between individuals as a payoff matrix game, we can model this as a shaking hand phenomena where \textbf{H} will sometimes pick the wrong action according to their own reasoning.
\item{\it A.I. systems can perform more frequent interactions:} Considering examples such as an A.I. support system for online shopping. An A.I. system can manage thousands of simultaneous auctions and so even if it gets a similar profit per transaction as a human, the larger number of transactions will provide greater gains. It can be modelled by allowing \textbf{A.I.} to play more frequently.
\item{\it A.I. systems can have access and analyze larger quantities of data:} With the current state-of-the-art there are many domains where people identify more complex relations, identify important variables and infer causality much better than machines \cite{ashton2015fly}. On the contrary, right now machines can analyze much larger volumes of data and variables that are already identified. Having more information about the world and specific problems combined with the ability to process such huge amounts of information can allow better prediction and consequently, better decision making capabilities. This advantage of A.I. systems is already being applied for stock exchange trading \cite{shen2012stock} and to create powerful language models \cite{brown2020language}. This advantage can be modelled by only giving \textbf{H} access to a noisy version of the matrix game. \textbf{A.I.} will be able to grasp the entirety of the problem and as such will have no noise in the observation of the matrix.
\end{enumerate}

There are other ways in which A.I. systems could grant an advantage to their users, some of which we might not even be able to understand yet given the current state of the technology. The main model assumption, however, remains the same: when interacting with \textbf{H}, \textbf{A.I.} have a decision making advantage. In the end, over the several different alternatives to model A.I. behaviours we choose item (3) to analyze in this work. In another simulation we use (1) and see that both alternative models provide similar qualitative conclusions. 

In a computational way we are in the presence of partial observability in our stochastic game. The difference between \textbf{H} and \textbf{A.I.} is that while \textbf{A.I.} sees the real payoff matrix, $M^t$, \textbf{H} sees a noisy version of it, $M^\epsilon$. This allows \textbf{A.I.} individuals to make optimal decisions, while \textbf{H} individuals are confined to sub-optimal decisions. This models the superior decision making and information gathering skills of \textbf{A.I.}. This approach also rests on the assumption that the individual value alignment problem is solved, since A.I. systems know the utility payoff of both individuals.

Having:
\begin{eqnarray}
(u_1^t, u_2^t) &=& M^t(a_1,a_2)
\end{eqnarray}
The noisy version is produced as follows:
\begin{eqnarray}
u_1^\epsilon &=& u_1^t + (z(0,10 - {Q}) - z(0,10 - {Q}))\\
u_2^\epsilon &=& u_2^t + (z(0,10 - {Q}) - z(0,10 - {Q}))
\end{eqnarray}

The degree of knowledge about the (true) payoff matrix $M^t$ is modelled in a continuous way. To do so, we consider a term $z(0,10 - {Q})$, where $Q$ corresponds to the level of intelligence. For $Q=10$ there is no noise and the true matrix is observed; $Q=0$ represents a low intelligence, such that the observed matrix is very different from the true one. \textbf{A.I.} is modelled with $Q=10$, while for \textbf{H} the intelligence factors is $Q\in[0,5]$. Other intervals for the intelligence factors of \textbf{H} were experimented with, inside the $[0,9]$ range, but they lead to the same qualitative results. The sum $(z(0,10 - Q) - z(0,10 - Q))$ was used instead of $z(-(10 - Q),10 - Q)$ to create a Irwin-Hall distribution instead of a uniform one.

As an example we can generate a $2$-by-$2$ true matrix (seen by \textbf{A.I.}) as:
\begin{eqnarray}
M^t=\left[\begin{array}[pos]{cc}
(0,0)  & (-3,1)\\
(1,-5) & (-1,-1)	
\end{array}
\right]
\end{eqnarray}
And then the noisy matrix observed by \textbf{H} becomes:
\begin{eqnarray}
M^\epsilon=\left[\begin{array}[pos]{cc}
(0,-1)  & (-1,3)\\
(0,-6) & (-2,1)	
\end{array}
\right]
\end{eqnarray}

Where each $(u_1^t, u_2^t)$ pair was transformed into the corresponding $(u_1^\epsilon, u_2^\epsilon)$ pair. In this example, $M^t(1,0)$ is $(1, -5)$ whereas $M^\epsilon(1,0)$ is $(0, -6)$. 
\subsection{Human Strategy:}
\label{sec:hum_strat}

Before delving into the different \textbf{A.I.} types, we describe the strategy used by \textbf{H}. Despite not having access to the true game matrix, $M^t$, \textbf{H} remain rational and will try to choose the actions most profitable for themselves. For this matrix game, that will correspond to the Nash equilibrium \cite{kalai1993rational,nash1951non,nash1950equilibrium}.

\subsubsection{\it Nash Equilibrium (NashEQ)} \textbf{H} play the Nash equilibrium in the noisy matrix $M^\epsilon$. If more than one is found, they choose the most profitable one. If two or more are equal, they choose the one most profitable for their opponent. If no Nash equilibrium is found, individuals choose the best action assuming that the opponent acts randomly.  

\subsection{A.I. Types:}
\label{sec:ai_strat}

In this section, we propose four different types of \textbf{A.I.}. A.I. systems can use the previously defined strategy for humans using the true matrix $M^t$ ({\it NashEQ}), but they can resort to more elaborate strategies ranging from a selfish to an utilitarian approach. \textbf{A.I.}, being modelled as having super-human intelligence, can also predict the action of an \textbf{H} opponent. \textbf{A.I.} cannot, however, predict opposing \textbf{A.I.} actions as for our model we assume all \textbf{A.I.} have equal intelligence and capabilities.

\subsubsection{\it Nash Equilibrium (NashEQ)} \textbf{A.I.} choose exactly like \textbf{H}, but using the true matrix $M^t$.

\subsubsection{\it Selfish}  \textbf{A.I.}, facing \textbf{H}, considers only its own profit, in accordance with ethical egoism \cite{rachels2012ethical}. Knowing what action \textbf{H} is going to take, \textbf{A.I.} chooses the action that maximizes its own payoff gain. When \textbf{A.I.} faces \textbf{A.I.}, they both choose according to the Nash Equilibrium method.

\subsubsection{\it Utilitarian} The other extreme is a pure utilitarian \cite{mill2016utilitarianism} A.I. system. \textbf{A.I.} facing \textbf{H} chooses the action that brings the greatest amount of payoff to the world, knowing what action \textbf{H} will take. This means that \textbf{A.I.} will choose the action that maximizes the sum between its own payoff and the payoff of \textbf{H}. When \textbf{A.I.} faces \textbf{A.I.}, it again chooses the action that maximizes the summed payoff of both players.

\subsubsection{\it Human Conscious (HConscious)} In between ethical egoism and utilitarianism, the objective of HConscious \textbf{A.I.} is to gather the greatest amount of payoff while, on average, avoiding negative impact on the \textbf{H} population. When \textbf{A.I.} faces \textbf{A.I.}, they both choose according to the Nash Equilibrium method. In practice, HConscious \textbf{A.I.} keeps two variables: $U$ that represents the summed payoff gain of all its previous \textbf{H} adversaries; and $E$, that represents the summed payoff those same \textbf{H} adversaries would have if they had faced a simulated \textbf{H}. When $U$ $\geq$ $E$, \textbf{A.I.} chooses an action that leads to a positive payoff to itself. When there are several such actions, the \textbf{A.I.} chooses the one that maximizes the utility payoff for the world, that is, that maximizes the sum of its own payoff and the opponent's payoff. If $U$ $<$ $E$, \textbf{A.I.} chooses an action that allows a positive payoff gain for its \textbf{H} opponent. Once again, when there are several such actions, the \textbf{A.I.} chooses the one that maximizes the utility for the world. 
Whenever the \textbf{A.I.} cannot find a positive action for himself (when $U$ $\geq$ $E$) or for its \textbf{H} opponent (when $U$ $<$ $E$), then it chooses according to the Utilitarian method.

\subsection{Simulation}
\label{sec:world}

We consider a world populated with $n$ individuals. $k$ of those are \textbf{A.I.} and the remaining $n-k$ are \textbf{H}, each having a randomly attributed intelligence, $Q$.

The fitness, or wealth, of an individual, $f$, be it either \textbf{H} or \textbf{A.I.}, is a measure of how well adapted it is to the world on which it is currently inserted. In our stochastic game model, the fitness of an individual is the sum of the payoff received after interacting (Sec.~\ref{sec:interaction}) once with all of the world's population of $n$ individuals.

For our simulations, the $n$ individuals that populate the world were set to interact randomly between each other over $N$ iterations. On each iteration, the following algorithm was used:
\begin{enumerate}
\item{Two individuals $I_1$ and $I_2$ are chosen at random from the population.}
\item{With probability $\mu=0.0005$, each individual can mutate and adopt an \textbf{A.I.} type or become \textbf{H}. In case of mutation return to step 1.}
\item{The fitness of $I_1$ and $I_2$, $f_1$ and $f_2$ respectively, is calculated.}
\item{If $I_1$ and $I_2$ are of different kinds (\textbf{A.I.} and \textbf{H}) or different \textbf{A.I.} types then $I_1$ imitates $I_2$ with a probability $p(f_1,f_2)$ (Sec.~\ref{sec:im_prob}).}
\item{If the imitation corresponds to adopting an \textbf{A.I.}, it can only do so if its fitness is above $P$, corresponding to the cost of buying a new A.I. system. Abandons (\textbf{A.I.} becoming \textbf{H}) and switching between \textbf{A.I.} types occur without any restrictions.}
\end{enumerate}

\subsection{Imitation Probability}
\label{sec:im_prob}

In our simulations, individuals can choose to adopt an A.I. system (\textbf{H} to \textbf{A.I.}) if they consider it advantageous, choose to abandon an A.I. system (\textbf{A.I.} to \textbf{H}), or change between \textbf{A.I.} types. Individuals may revise their choices through social learning. For instance, an \textbf{H} can decide to imitate an \textbf{A.I.} following a Selfish choice behaviour if it finds such \textbf{A.I.} has a significantly better fitness than its own. On such imitation, the individual would stop being \textbf{H} and become \textbf{A.I.}.

Using this idea, let us now detail how a population of self-regarding individuals revise their choices. At each time-step an individual $x$ is randomly selected to revise its choices. This individual will imitate a randomly chosen individual $y$, with a probability ${p}({f_x},{f_y})$, that increases with the fitness difference between $y$ and $x$, given by $f_x$ and $f_y$, respectively. Here we adopt the Fermi update or pairwise comparison rule \cite{traulsen2006stochastic}, commonly used in the context of evolutionary game theory and population dynamics in finite populations \cite{sigmund2010calculus}, where $p$ is given by
\begin{equation}
    {p}({f_x}, {f_y}) = \frac{1}{1 + e^{-\beta({f_y}-{f_x})}}
\end{equation}
in which $\beta$ translates the noise associated with the imitation process. Throughout the simulations we have $\beta = 0.1$. As a result of this process, the strategy of individuals with higher fitness will tend to be imitated, and spread in the population.

\section{Results}
\label{sec:results}

We will now study the properties of the stochastic game in terms of equilibrium points between the different types of population and their relative fitness. This will give us insights into the adoption and acceptance of A.I. systems. Given the inherent stochastic nature of our model, all the presented results are averaged over 20 runs.

\subsection{A.I. systems' adoption}
\label{sec:adoption}

In this initial results we will answer the two first questions.
To do so, we perform a simulation where we allow all types of A.I. systems (presented in Sec.~\ref{sec:ai_strat}) to compete to be adopted by humans. The initial condition is $90\%$ of \textbf{H} and $10\%$ distributed uniformly among 4 types of \textbf{A.I.}. We let the system run for $20,000$ iterations.  

\begin{figure*}[!htb]
\minipage{0.48\textwidth}
  \includegraphics[width=\linewidth]{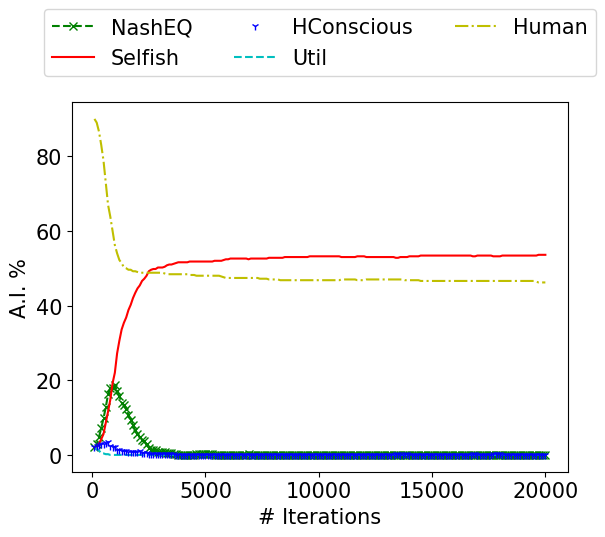}
  \subcaption{}\label{fig:ai_it_37_cost}
\endminipage\hfill
\minipage{0.48\textwidth}

    \includegraphics[width=\linewidth]{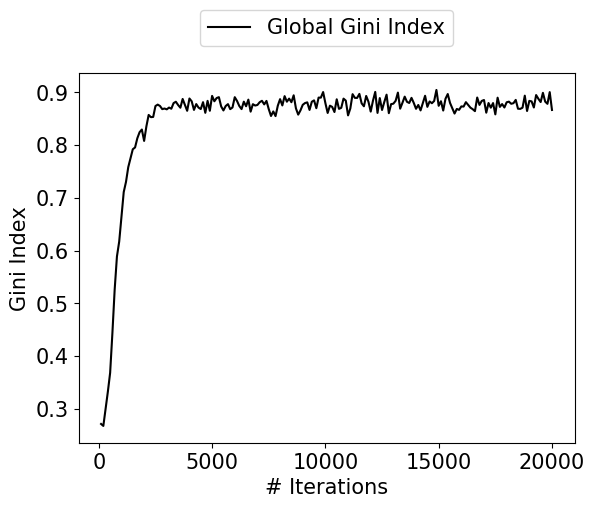}
  \subcaption{}\label{fig:gini_it_37_cost}
 
\endminipage\hfill
\minipage{0.02\textwidth}
    \rotatebox{90}{\space \space \space \space \space \space inequality increases\\$\longrightarrow$}
\endminipage\hfill
\caption{Evolution of the \% of \textbf{A.I.} (a) and Gini coefficient (b) in a world with an \textbf{A.I.} cost of 37 $(P = 37)$, showing a world that becomes populated with $54\%$ Selfish \textbf{A.I.} and 46\% \textbf{H} (a), having a high inequality (Gini $\approx 0.88$) (b). For this simulation we have $N = 20000$, $n = 500$ and $P = 37$ was chosen as it corresponds to $25\%$ of the average utility gathered by a \textbf{H} only population of $500$. Other values were tested, but they led to the same qualitative results. The \textbf{H} population can be seen quickly becoming Selfish \textbf{A.I.} up to a point where the remaining \textbf{H} don't have enough fitness to become \textbf{A.I.}. This leads to an equilibrium where high fitness Selfish \textbf{A.I.} take advantage of low fitness \textbf{H}, giving rise to a high degree of inequality in the world.}
\end{figure*}

The evolution of the percentage of population that adopted each type of A.I. system is shown in Fig.~\ref{fig:ai_it_37_cost}. We can observe a final equilibrium where $54\%$ of the population became Selfish \textbf{A.I.}, and $46\%$ did not adopt any A.I. system. At this equilibrium, the average fitness for Selfish \textbf{A.I.} is $370$, whereas the fitness for non-adopters \textbf{H} is $-104$, representing a very unequal society. This is confirmed in Fig.~\ref{fig:gini_it_37_cost} where we portray the evolution in time of the Gini coefficient, a measure of inequality in the population (Appendix~\ref{sec:gini}.) for the entire population. To be used as a baseline we compute the fitness of an all \textbf{H} population (Table~\ref{tab:avgdata}). The fitness in this case is 149 with a Gini coefficient of 0.17. Overall, the non-adopters become much worse than they would be in a world without A.I. systems while adopters become much better. 

This equilibrium can be understood in an intuitive way. Early adopters are able to gather fitness much faster so that latter adopters cannot meet the buying price for A.I. systems. This explains the co-existence between adopters and non-adopters. Similar results are obtained for other cost values, $P$, noting that the higher the value of $P$, the lower the \% of \textbf{A.I.} in the final equilibrium. 
Overall, this simulation shows that people have an incentive to adopt an A.I. system, albeit a Selfish one. As a result, we observe the emergence of an unequal society where adopters largely increased their fitness while non-adopters lost their fitness.

\subsection{Rationality of adoption and acceptance}

Here we will answer the third research question.
The previous section showed a non-trivial relation between AI adoption and the particular strategy artificial systems have. To better understand such emerging dynamics, in this section we describe the characteristic dynamics created by each type of \textbf{A.I.}.

Table~\ref{tab:avgdata} shows the average fitness and wealth inequality obtained in the case of a homogeneous society of \textbf{H}, and each type of \textbf{A.I.}. It suggests that Utilitarian \textbf{A.I.} would provide the best overall fitness with less inequality. All the other \textbf{A.I.} strategies are shown to provide similar fitness values to a homogeneous \textbf{H} society without A.I. systems. The lower values in the Gini coefficient are due to the reduction in noise due to the perfect observability of \textbf{A.I.}, and the different intelligence values, $Q$, between \textbf{H}.
\begin{table}[!htbp]
    \centering
    \caption{Average fitness and Gini coefficient on a world fully populated by \textbf{H} or by \textbf{A.I.} following a single behaviour. Except for the utilitarian, the differences in fitness are not statistically significant. }
    \label{tab:avgdata}
    \begin{tabular}{lll}
    \toprule
    \midrule
                        & fitness& Gini\\\hline
    Human (100\%)       & 149    & 0.17\\
    NashEQ (100\%)      & 150    & 0.14\\
    Selfish (100\%)     & 150    & 0.14\\
    HConscious (100\%)  & 149    & 0.14\\
    Util (100\%)        & 378   & 0.08\\
    \bottomrule
    \end{tabular}
\end{table}

A world fully populated by Utilitarian \textbf{A.I.} would be better for everyone. However, we know that if other types of A.I. systems are present, the Selfish behaviour prevails and the Utilitarian is abandoned (Sec.~\ref{sec:adoption}). This will naturally have an impact in the dynamics of adoption of A.I. systems.

\begin{figure}[!htb]
    \centering
    \includegraphics[width=0.48\textwidth]{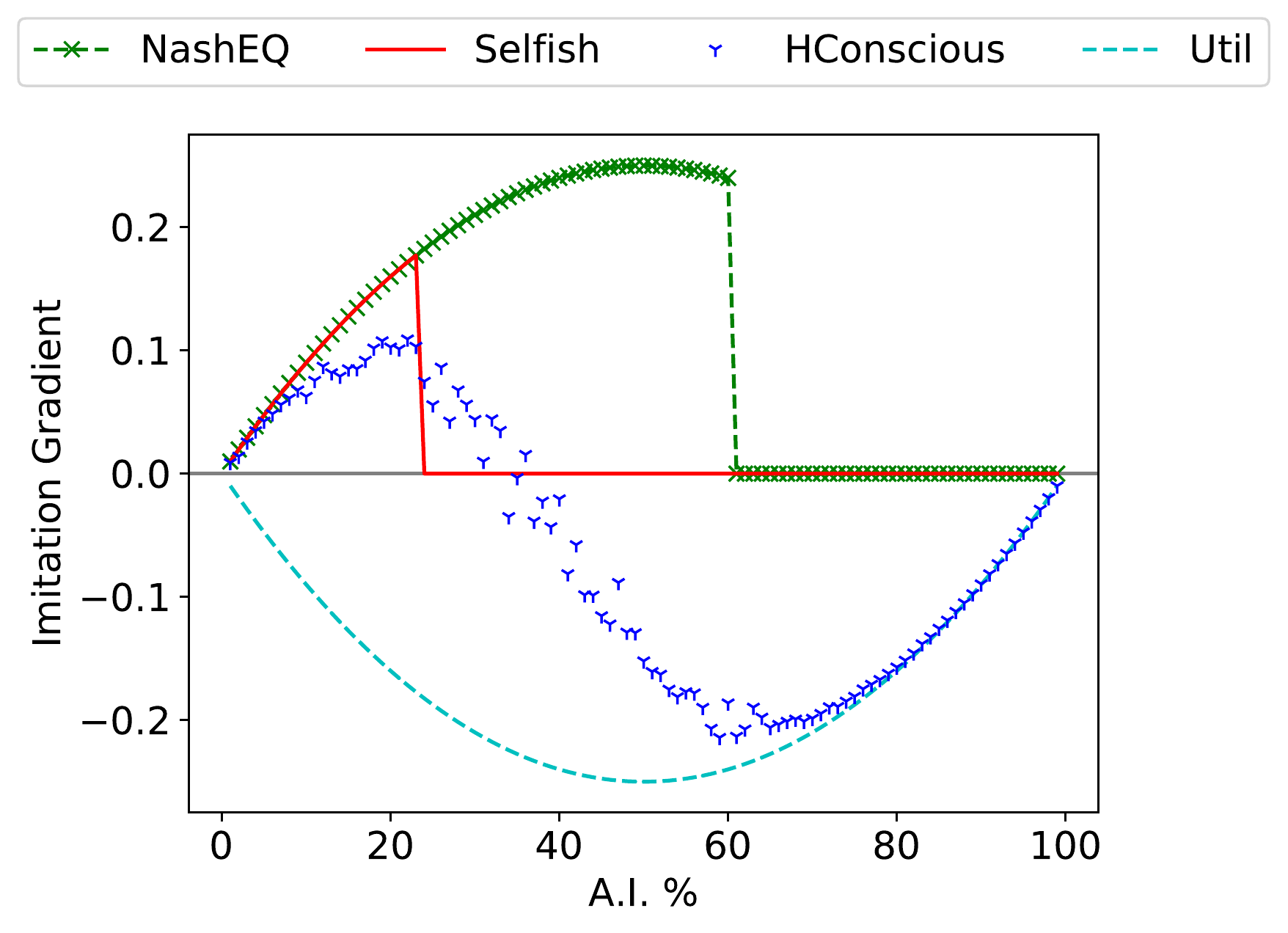}
    \caption{Imitation gradient plots for the different choice behaviours in a world with an \textbf{AI} cost of 37 $(P = 37)$, $n=500$. {Positive gradients mean that on average, the number of \textbf{H} wanting to adopt an A.I. system will tend to increase; negative gradients mean that, on average, \textbf{A.I.} adoption is likely to decrease. The sharp drops on both the Selfish and NashEQ behaviours correspond to the equilibrium point where the \textbf{H} population can no longer afford to become \textbf{A.I.}. The Utilitarian behaviour is never freely adopted and Utilitarian individuals always prefer to become \textbf{H}. The HConscious behaviour shows the most interesting dynamic, having either a positive or a negative imitation gradient depending on the \% of \textbf{A.I.} individuals. A population of free choosing \textbf{H} and HConscious \textbf{A.I.} will maintain a stable equilibrium at around 40\% HConscious \textbf{A.I.}.}}\label{fig:eq_plot_37Cost}
\end{figure}

In Fig.~\ref{fig:eq_plot_37Cost} we show the imitation gradient $G$ as a function of the fraction of \textbf{A.I.}, for different \textbf{A.I.} types. A description of how the imitation gradient is calculated can be found on Appendix~\ref{sec:im_grad}. Whenever $G>0$ ($G<0$) the fraction of A.I. will tend to increase (decrease). We can see that, depending on the type of \textbf{A.I.}, different dynamics and equilibrium points emerge (Table~\ref{tab:avgdata_equi}).  The Utilitarian strategy is always disadvantageous and is unlikely to be adopted by \textbf{H}. Differently, NashEq, Selfish and HConscious strategies favour the co-existence of \textbf{H} and \textbf{A.I.}.

\begin{table*}[!htbp]
    \centering
    \caption{Average \textbf{H} fitness, average \textbf{A.I.} fitness, average total fitness and Gini on a world in the equilibrium point of each type of \textbf{A.I.} and \textbf{H}. Util $100\%$ is a top baseline but it is not an equilibria point. In parenthesis we show the ratio to an $100\%$ population of \textbf{H}. Of the equilibria, only the HConcious behaviour provides an advantage for both \textbf{H} and \textbf{A.I.}.}
    \label{tab:avgdata_equi}
    \begin{tabular}{rlcccc}
    \toprule
    \midrule
    &               & \textbf{H} & \textbf{A.I.} & Total & Gini\\\hline
    \multirow{4}{*}{\rotatebox{90}{Equilibria\space \space \space \space \space \space \space \space}} 
    &Human (100\%)      & 149            & - & 149 & 0.17\\\hline
    &NashEQ (60\%)      & 38(0.25$\downarrow$)& 229(1.53$\uparrow$) & 152(1.02$\uparrow$) & 0.38(2.24$\downarrow$)\\
    &Selfish (25\%)     & 31(0.21$\downarrow$)& 510(3.42$\uparrow$) & 151(1.01$\uparrow$) & 0.70(4.12$\downarrow$)\\
    &HConscious (40\%)  & 172(1.15$\uparrow$) & 168(1.23$\uparrow$) & 170(1.14$\uparrow$) & 0.15(0.88$\uparrow$)\\
    &Util (0\%)         & 149(1.00) & - & 149(1.00) & 0.17(1.00)\\\hline
    &Util (100\%)   & - & 378(2.54$\uparrow$) & 378(2.54$\uparrow$) & 0.08(0.47$\uparrow$)\\\hline
    \bottomrule
    \end{tabular}
\end{table*}

We observe that the best equilibrium for society in general is the HConscious (40\%), having a low Gini coefficient of 0.15 and an improved utility values for both \textbf{H} and \textbf{A.I.} compared to the Human (100\%) baseline. Both the NashEQ(60\%) and the Selfish(25\%) equilibria improve the utility of the \textbf{A.I.} population at the cost of the \textbf{H} population, leading to an increase in inequality (higher Gini coefficient). 

We find that all equilibria are worse for society than the fully Utilitarian \textbf{A.I.} population. We also note that for the equilibrium shown in Fig.~\ref{fig:ai_it_37_cost} the average fitness of the Selfish \textbf{A.I.} population is 370, which is less than the obtained by the \textbf{A.I.} population at Util ($100\%$). However, at Selfish ($25\%$), the average fitness of the Selfish \textbf{A.I.} population is 510, greater than both the previously mentioned fitness. We note that the equilibrium between \textbf{H} and \textbf{A.I.} using exclusively a Selfish A.I. system is different from the one observed in Fig.~\ref{fig:ai_it_37_cost}. The presence of other \textbf{A.I.} types in the population allowed a greater number of individuals to afford a Selfish A.I. system compared to a world where only \textbf{H} and Selfish \textbf{A.I.} are present.

\begin{figure}[!htb]
    \centering
    \includegraphics[width=0.45\textwidth]{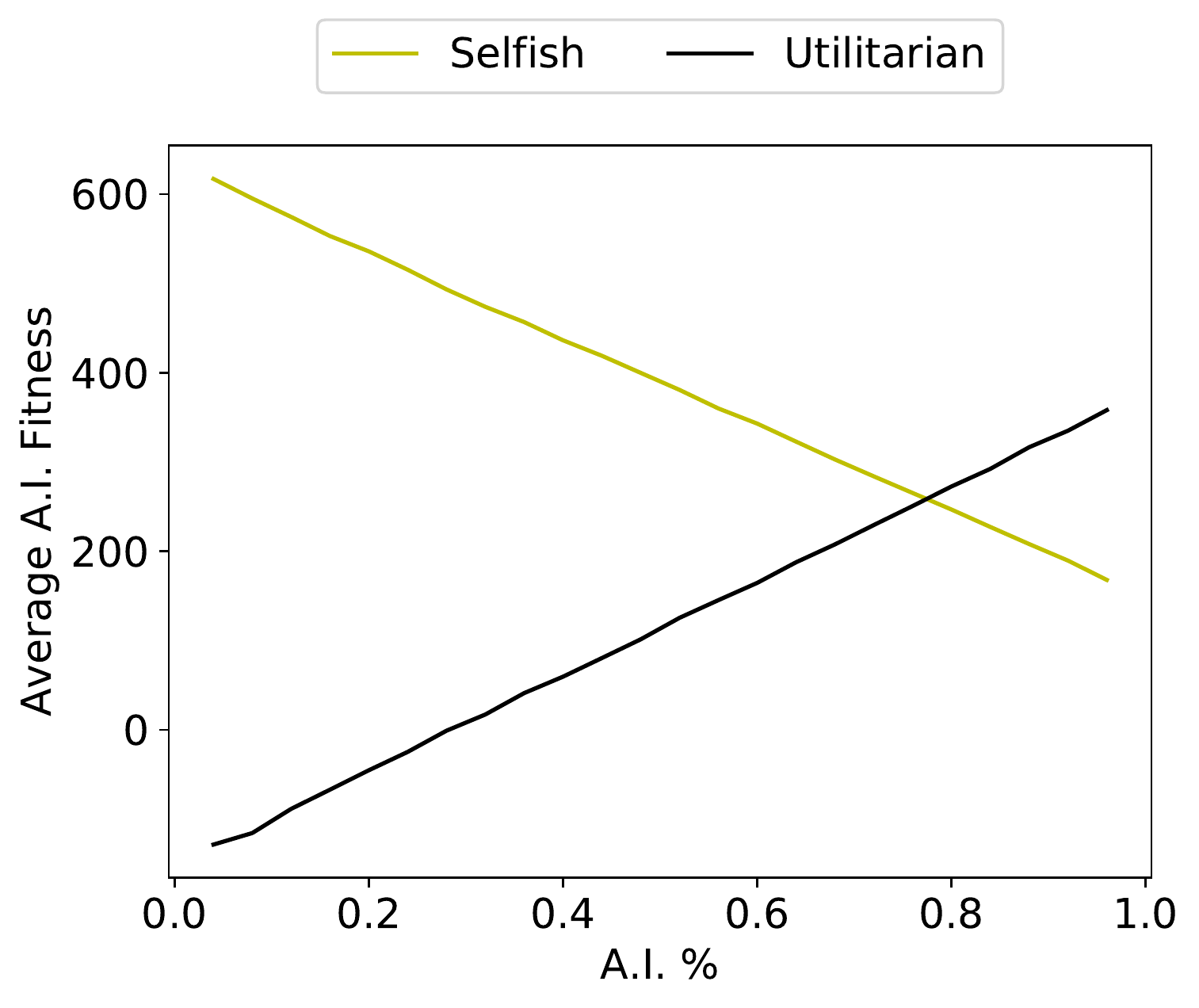}
    \caption{\textbf{A.I.} average fitness evolution with the percentage of \textbf{A.I.} population, showing the contrasting fitness dynamics between the Utilitarian and the Selfish behaviours. Whereas Utilitarian \textbf{A.I.} benefits from a higher percentage of \textbf{A.I.} population, Selfish \textbf{A.I.} incurs a loss of fitness. In a population with more than 80\% of \textbf{A.I.} individuals, it becomes more beneficial, even for the \textbf{A.I.} individuals, to have all \textbf{A.I.} be Utilitarian.}
    \label{fig:great_self}
\end{figure}

Studying the relation between the percentage of the \textbf{A.I.} population and the average fitness for the Selfish and Utilitarian \textbf{A.I.}, we found that the fitness for Selfish \textbf{A.I.} reduces the greater the \% of \textbf{A.I.} in the population whereas the opposite occurs with the Utilitarian \textbf{A.I.} (Fig.~\ref{fig:great_self}). This behavior shows that having a great part of the population using a Selfish A.I. system ($<78\%$) is worse even from the individual point of view of A.I. adopters, that would be better off if they were Utilitarian \textbf{A.I.}.

We also note that the cost of adopting an A.I. system has an effect on the final equilibrium. Experimenting with several different values of P we found that the higher the cost of adoption, the lower the \% of \textbf{A.I.} on the population and consequently, the higher the difference in fitness between \textbf{A.I.} and \textbf{H}. The same effect was found when we lowered the value of the matrix inflation, $\alpha$.

\begin{figure*}[!htb]
\minipage{0.48\textwidth}
  \includegraphics[width=\linewidth]{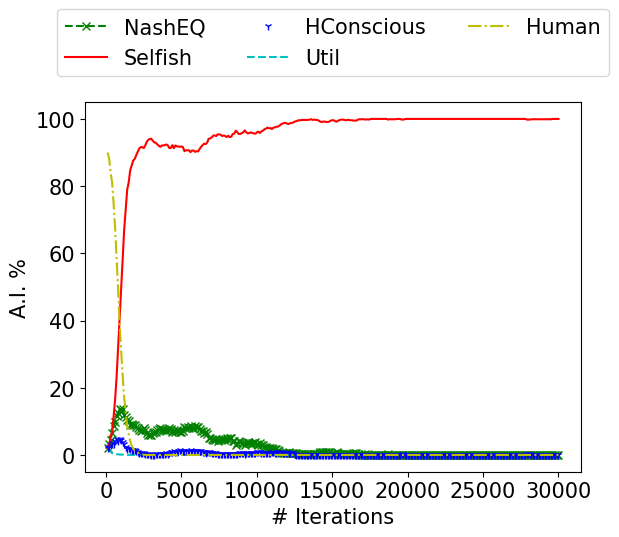}
  \subcaption{}\label{fig:num_ai_no_cost}
\endminipage\hfill
\minipage{0.48\textwidth}
    \includegraphics[width=\linewidth]{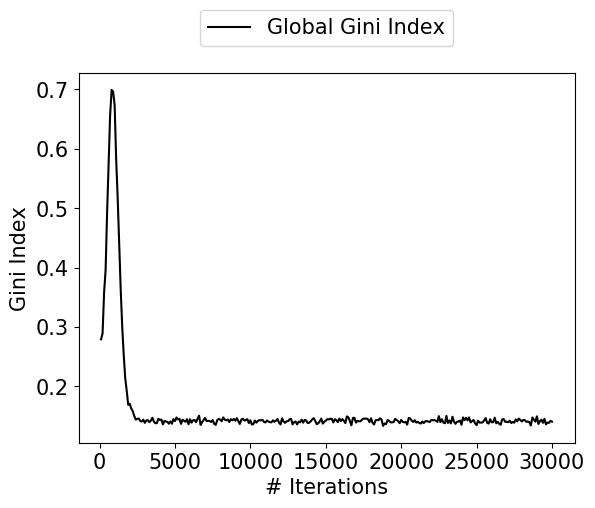}
  \subcaption{}\label{fig:gini_no_cost}
\endminipage\hfill
\minipage{0.02\textwidth}
    \rotatebox{90}{\space \space \space \space \space \space  inequality increases\\$\longrightarrow$}
\endminipage\hfill
\caption{Evolution of the \% of \textbf{A.I.} (a) and Gini coefficient (b) in a world with no \textbf{A.I.} cost $(P = -\infty)$, showing a world that becomes populated with a mostly Selfish \textbf{A.I.} population (89\%) (a) but stabilizing with a low inequality (b). For this simulation we have $n = 500$ and $N = 30000$. Without the adoption cost constraint, there is nothing preventing all \textbf{H} from becoming \textbf{A.I.}. The Gini index dynamics sharply contrast those of Fig. 1. There is a spike in inequality but, as all \textbf{H} are able to become \textbf{A.I.}, inequality quickly decreases and stabilizes at a low level. 
}
\end{figure*}

\subsection{Dynamics of adoption in cost free A.I.}

Here we study a variant of the previous simulations, where we considered a cost for adopting an A.I. system. We decided to study what would happen if there was no such cost, that is, if anyone could freely adopt an A.I. system regardless of its current fitness. In practical terms, this meant setting $P = -\infty$.

When all \textbf{A.I.} types co-exist, the absence of a cost significantly change the dynamics. As there is no fitness barrier to becoming \textbf{A.I.}, the entire population does become \textbf{A.I.}. Once again the Selfish behaviour dominates the population after around $2\times10^4$ iterations. After $2.5\times10^3$ iterations, there are no more \textbf{H} present on the population\footnote{Different \textbf{A.I.} types do not have any particular advantage when playing against other \textbf{A.I.}, but the Selfish population does have an advantage as soon as a \textbf{H} appears in the population thanks to a mutation.}. This leads to an increase in the number of Selfish \textbf{A.I.} every time such a mutation occurs. That steady increase lasts until the mutated \textbf{H} imitates an \textbf{A.I.} type. This leads to a fully Selfish world instead of a world where several \textbf{A.I.} types coexist. Regarding the Gini coefficient, it stabilizes, after a sharp peak, around a relatively low value ($\approx 0.14$).
\begin{figure}[!htb]
    \centering
  \includegraphics[width=0.48\textwidth]{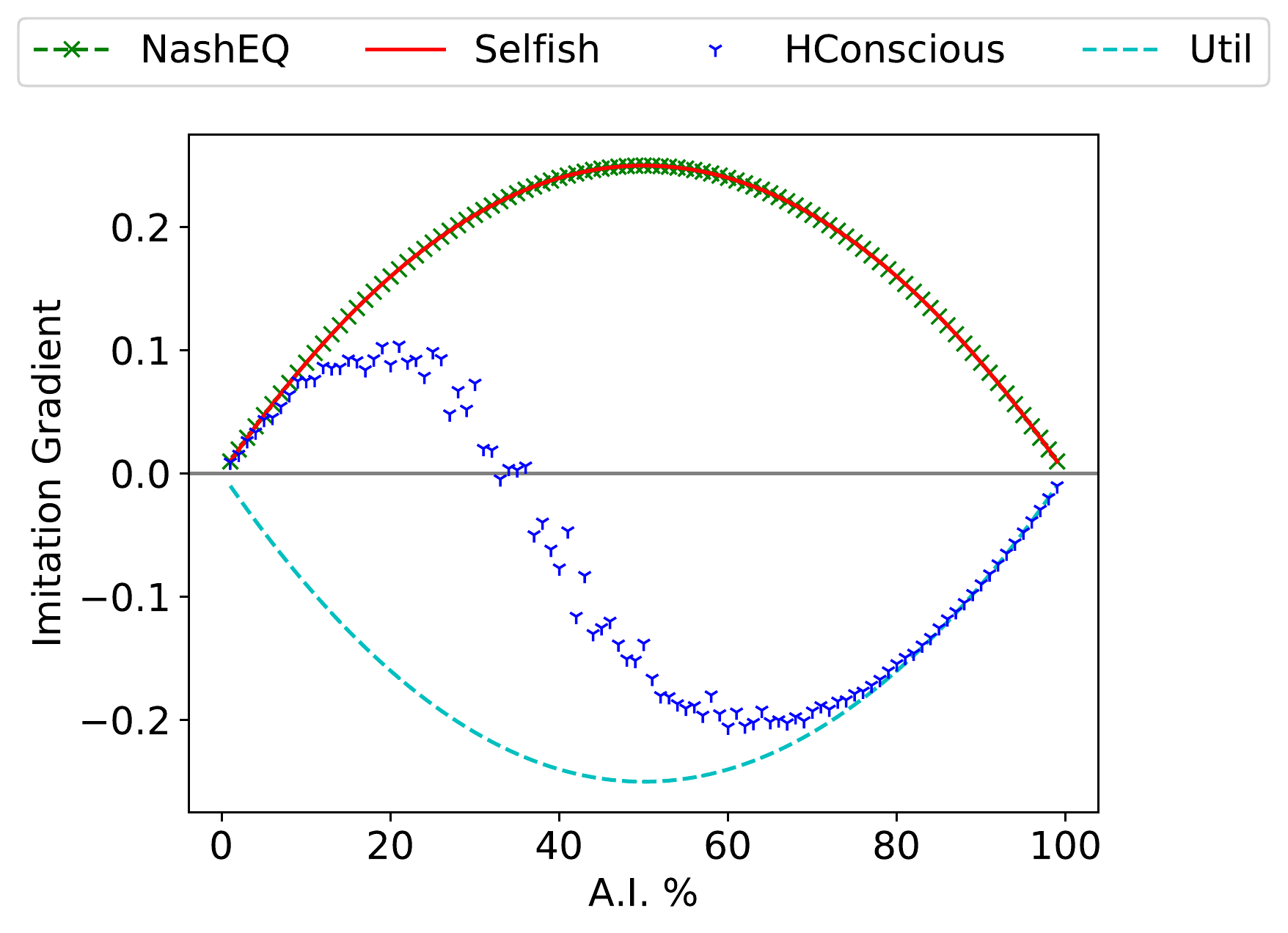}
  \caption{Equilibrium plots for the different choice behaviours in a world with no A.I. system adoption cost $(P = - \infty)$ and $n$ = 500. {Positive gradients mean that on average, the number of \textbf{H} wanting to adopt an A.I. system will tend to increase; negative gradients mean that, on average, \textbf{A.I.} adoption is likely to decrease. The sharp drops present on Fig. 2 on both the Selfish and NashEQ behaviours are no longer present without the adoption cost constraint. The Utilitarian behaviour continues to never be freely adopted and Utilitarian individuals always prefer to become \textbf{H}. The HConscious behaviour mantains the same dynamic, having either a positive or a negative imitation gradient depending on the \% of \textbf{A.I.} individuals. A population of free choosing \textbf{H} and HConscious \textbf{A.I.} will maintain a stable equilibrium at around 40\% HConscious \textbf{A.I.}.}}
  \label{fig:eq_plot_noCost}
\end{figure}
The impact of a cost-free AI, can be easily understood looking at the imitation gradient for a population of \textbf{H} and \textbf{A.I.} (Fig.~\ref{fig:eq_plot_noCost}). In this case, the imitation gradient no longer goes to 0 in the Selfish and NashEQ behaviours, as there is no cost to limit the imitations. The new equilibria are:
\begin{enumerate}
    \item Human(0\%) / NashEQ(100\%)
    \item Human(0\%) / Selfish(100\%)
    \item Human(60\%) / HConscious(40\%)
    \item Human(100\%) / Util(0\%)
\end{enumerate}

What we find for a world without an A.I. system adoption cost is that the equilibria tend to lead to more egalitarian worlds. When all choice behaviours are present in the population, we end up with a fully \textbf{A.I.} population (Fig.~\ref{fig:num_ai_no_cost}), which leads to an average total utility of 150, around the same as if we had a fully \textbf{H} population, and to a Gini coefficient of $\approx0.14$, lower than with the fully \textbf{H} population. Does this mean that as long as there is no significant cost to the adoption of an A.I. system, the world will remain the same or even improve in terms of equality? We explore this in the following section.

\input{asym.tex}

\section{Conclusion}
\label{sec:conclusions}

In this work, we study the adoption, acceptance, and impact on the individual and societal  fitness (including the disparity of fitness measured with the Gini coefficient) of A.I. systems that work as a proxy for humans. To do so, we developed a stochastic game theoretical model to simulate \textbf{H} and \textbf{A.I.} interactions. 

We began this paper with 3 research questions, which we are now able to answer with regards to our simulated environment. (1) Will self-regarding individuals adopt A.I. systems? Yes, they will, both when there is an adoption cost (Fig.~\ref{fig:ai_it_37_cost} and Fig.~\ref{fig:eq_plot_37Cost}) and when there is no cost of adoption (Fig.~\ref{fig:num_ai_no_cost} and Fig.~\ref{fig:eq_plot_noCost}).
(2) With different types of A.I. systems available, which ones will be adopted? When all types of A.I. systems are available, the one that is predominantly adopted is the Selfish one (Fig.~\ref{fig:ai_it_37_cost} and Fig.~\ref{fig:num_ai_no_cost})
(3) If adopted, what is the individual and collective gain, depending on the strategy adopted by the A.I. system? The answer to this question can be found on Tab.~\ref{tab:avgdata} for worlds that are fully populated by a single type of \textbf{A.I.} and in Tab.~\ref{tab:avgdata_equi} for the equilibrium states between \textbf{H} and a single \textbf{A.I.} type.

Our main conclusion is that without regulation and considering an A.I. system adoption cost, pure selfish A.I. systems will be adopted by a part of the society until those early adopters accumulate all fitness to the point that non-adopters are unable to adopt an A.I. system. As a result, A.I. adopters have a significant increase in fitness while the remaining population will be much worse off than in a world without A.I. systems. This leads to an unequal society (high Gini coefficient), and as such there is a high probability non-adopters will not accept the existence of such A.I. systems.

Analyzing each type of A.I. system independently, we can see that a world entirely populated by Utilitarian \textbf{A.I.} would be the best for society. However, that type of A.I. system is not individually rational, and, as such, a world entirely populated by Utilitarian \textbf{A.I.} can be easily exploited by Selfish \textbf{A.I.} or even by \textbf{H} and will never prevail on an emerging equilibrium.

When allowing only one single \textbf{A.I.} type in the world, the HConscious type of \textbf{A.I.} displayed an interesting property: there is an equilibrium point (at around $40\%$ of \textbf{A.I.}) where \textbf{A.I.} co-exist with \textbf{H} resulting in {\it i)} an increase in fitness for adopters and non-adopters, and {\it ii)} a reduction in inequality (lower Gini values)(Fig.~\ref{tab:avgdata_equi}). Here, we claim that even non-adopters will accept the existence of A.I. systems as they also obtain a gain.  This means that if Human Conscious A.I. is the only norm available, it will be adopted up to a certain equilibrium, resulting in an overall gain for both \textbf{A.I.} and \textbf{H}. Comparatively, the other \textbf{A.I.} types led to either prejudicial equilibria for the \textbf{H} population (NashEQ and Selfish) or to the $100\%$ \textbf{H} equilibrium (Fig.~\ref{tab:avgdata_equi}).

When studying the case of cost-free adoption of A.I. systems, we observed that in a world with all \textbf{A.I.} types available, the final equilibrium is a mostly Selfish \textbf{A.I.} population. Unlike the previous simulations with an adoption cost, this final equilibrium does not create a societal gap and leads to a slight decrease in the Gini coefficient. The average fitness of the population also remains the same. This is not an adverse outcome but leaves us far away from the optimal result we can obtain with the thoroughly Utilitarian \textbf{A.I.} population and worse than with the equilibrium of HConscious (40\%).

Once again, analyzing each \textbf{A.I.} type individually, we notice that only Human Conscious \textbf{A.I.} allows us to reach an equilibrium that improves upon the baseline of an all \textbf{H} world. This result furthermore consolidates that, if only one \textbf{A.I.} type was to be available in a free choice society, it should be this one. 

The Utilitarian behaviour is the one that allows us to reach the maximum fitness of all the ones here studied, but being easily exploited, it isn't short term individual rational and as such, not adopted. Even if all the population was initially Utilitarian \textbf{A.I.}, the appearance of a single \textbf{H} could result in the entire \textbf{A.I.} population abandoning the Utilitarian A.I. system and choosing to become \textbf{H}.

Finally, we analyzed the impact of A.I. systems in a fully \textbf{A.I.} populated society where some part of the population behaves in a parochial/discriminatory way. We observed that individuals belonging to a majority will have more significant gains than people in minorities and that more powerful A.I. systems exacerbate this effect. For instance, if some part of the population has some discriminatory behavior, when equipping the society with A.I. systems that are more efficient, the impact of anti-social behavior will be stronger.

\section{Discussion}
\label{sec:discussion}

We developed this work to understand the dynamics of adoption and acceptance of A.I. systems, in a world where individuals prefer what offers them an advantage. This \textit{individual rationality} will be important for the survival of A.I. systems, their commercial success and research investment. Under the assumptions of our model, any advantage provided by A.I. systems will lead to their adoption without the need for any regulation.

Unfortunately, there will be types of A.I. systems that will exploit human weaknesses, such that the profit for their owners will be at the cost of non-adopters of such technology. This situation fails at respecting the \textit{societal rationality}. In this case, people that cannot afford, or do not have access to such A.I. systems, might push an agenda of strong regulation against them.

There are nevertheless some types of A.I. systems that assume norms that help their users without exploiting others, increasing the individual and the overall fitness of the society while reducing inequality. These types of Human Conscious A.I. systems have a dynamic that makes them be adopted until they reach a stable equilibrium with non-adopters. Differently, a purely utilitarian A.I. system could provide even more gains to society but would not reach a stable state, as \textbf{H} could easily exploit Utilitarian \textbf{A.I.}. Having everyone relying on purely utilitarian A.I. systems could only be achieved with a strong regulation/coercion. 

When we remove the cost of adoption of A.I. systems, we reach an entirely Selfish \textbf{A.I.} society. This society presents an equal average fitness to a fully \textbf{H} society and a lower Gini coefficient. At first glace, this seems like a positive outcome. Not the best we could achieve, but one that requires no regulation. However, as we show in Sec.~\ref{sec:asym}, such a world could lead to an increase in inequality as soon as, for instance, we introduce parochial individuals. It is not enough to remain mostly the same when A.I. systems are introduced. There needs to exist a significant improvement for society as a whole in order to compensate for the inequality that arises thanks to the exacerbation of the effects of the already present human parochialism. Either that or we must ensure that A.I. systems are unable to behave in a parochial way, even if their users do so.

Furthermore, we can expect that having a competitive A.I. system will always come at some sort of cost. Even if with time, the price of having a relatively smart system becomes negligible, it isn't far fetched to assume that having a top of the line system won't be readily available to everyone unless strong regulations are in place. We can thus claim that some regulation is needed. The only way to ensure that an A.I. system is both individual and societal rational, is to enforce that the A.I. system is Human Conscious, that is, beneficial for their owners without harming non-adopters. Such regulations shouldn't be restricted to frequent assessments on whether A.I. systems are having or not a Human Conscious impact. To ensure a truly robust system and following the ideas of Yampolskiy, regulations should require strict proof that the A.I. system can't help but be Human Conscious, even under recursive self-improvement \cite{yampolskiy2013safety}.

Other regulation might be needed. As previously discussed, to avoid that A.I. systems exacerbate the antisocial behaviour of parts of the population, it might be required that A.I. system cannot align to the parochialism of their users. We can also require this due to an ethical perspective, avoiding prejudice, discrimination, and lack of tolerance towards minorities. Another perspective of this rule is to ensure that adopters of A.I. systems do not collude with each other to further increase their gains. For instance, owners of autonomous vehicles could agree between themselves that in case of an accident they would favour crashing into a non-owner of an autonomous vehicle than to a fellow autonomous vehicle owner.

Pushing for Human Conscious \textbf{A.I.} and non-discrimination can be achieved in different ways. We could approach the problem with a purely legislative perspective, but we could also partially rely on cooperation induction mechanisms \cite{rand2013human,sigmund2010calculus}. For example, in worlds where individuals have reputations, we could stipulate that individuals that choose to use a utilitarian approach gain a positive reputation \cite{nowak2005irreview,santos2018nature}. Individuals could then choose only to interact with those who have a positive reputation, giving an extra incentive for the adoption of Utilitarian or Human Conscious A.I. systems. Moreover, decision-making in networked populations could result in clusters of Utilitarian \textbf{A.I.} that could be stable as they mostly interact between themselves \cite{nowak1992evolutionary,rand2014static}. Also, the heterogeneous nature of human interactions may further increase the chances of reaching pro-social \textbf{A.I.} \cite{santos2006pnas,santos2008nature}. It could also introduce a natural form of parochialism, as clusters would mostly interact between themselves.

In this work, we assume individuals are fundamentally selfish, only adopting an A.I. system if it is in their own self interest to do so. However, it might be argued that individuals might decide to adopt an A.I. system if they consider it beneficial for the world, even if not directly beneficial for themselves. Trying to model this altruistic adoption could be a topic for future work \cite{paiva2018engineering}.

There are still some weaknesses in just ensuring that non-adopters are not worse than before. If a part of the population gets better while another part of the population stays the same, in practice, they become worse in relative terms. This is already the case with access to education and health: no one is worse off by having other people going to the doctor or school in comparison to no one having access to this services, but they are worse in relative terms. However, if the living conditions of non-adopters increase due to new and more efficient services being available, then the total fitness might have improved and thus the fitness of each individual.

One starting assumption in our work was that the individual value alignment problem was correctly solved. This assumption was modelled in the perfect observation by A.I. systems of the payoff matrix for both their owners as for the \textbf{H} they interacted with. However, learning human values is not trivial. A machine looking at the behaviour of people might assume that the observed behaviours correspond to what people think is ethical. Unfortunately, examples abound where people's actions do not correspond to their preferences or ethics. Looking at extreme behaviours, we can consider slavery as an example. An A.I. system might assume that slaves choose to work instead of doing nothing. Of course, the explanation, in this case, is an unobserved variable: the not working action would correspond to a much higher penalty than working under duress. In everyday life, similar examples occur where hidden variables are required to explain why people are making choices that seem to go against their best interests. Other simple cases occur when people pay higher prices for single items where a pack of multiple items would give them a discount, here a non-observed constraint on the available money to invest would also explain such behaviours. Creating simulations where the \textbf{A.I.} has to learn the values of their owners and then act accordingly could provide many insights, and is a strong avenue for future work. We note that when people have to program an A.I. system explicitly, they behave in a fairer way than when acting directly \cite{de2018social}.

Any multi-agent system model of human societies is always an abstraction of infinitely more complex behaviours between people. We can think of other types of interactions for which the same behaviours would have different properties. One avenue of interest could be to combine a matrix game with an ultimatum type of game. In this setting after observing the payoffs any of the individuals could refuse the deal. The individual better informed might have a higher incentive to propose a fair deal to avoid rejection. In a more closely related setting with the ultimatum game where only one individual chooses the action, and the other can refuse or not again the balance of powers would shift. Another limitation of our model is that we collapsed all interactions between individuals to a one-time choice of a single action requiring no long-term planning. We could also assume a cost model for becoming \textbf{A.I.} as a pay per interaction and not a lump sum at adoption time. This would change the dynamics as more people could become \textbf{A.I.}. This could prove particularly interesting in the cases where the extra capabilities of the A.I. system are used to help others such as in the Human Conscious behaviour.

\begin{appendix}

\section{Gini coefficient}
\label{sec:gini}

In order to compare the emerging inequality between different runs of our simulations and to understand how the inequality varies within each simulation, we'll calculate the Gini coefficient based on the fitness of the population. The Gini coefficient is a measure of statistical dispersion, being the most commonly used measurement of inequality. In economics, it is often used to assess the income disparity inside a given country. A Gini coefficient of 0 expresses perfect equality, where everyone has the same fitness, whereas a Gini coefficient of 1 expresses maximal inequality.

Several different approaches to calculate the Gini coefficient have been proposed. Based on the fact that the Gini coefficient is half the relative mean absolute difference and that the relative mean absolute difference is the mean absolute difference divided by the arithmetic mean, for our simulations, we use the following definition:
\begin{multline}
Gini(f_1,...,f_n) =\\ \frac{1}{2}\Bigg( \frac{\frac{1}{n^2}\sum_{i=1}^{n}\sum_{j=1}^{n}|f_i-f_j|}{\frac{1}{n}\sum_{k=1}^{n}f_k}\Bigg)
\end{multline}

Where $f_1,f_2,...,f_n$ correspond to the fitness of each individual.

\section{Imitation gradient}
\label{sec:im_grad}

To better understand the desire of the \textbf{H} population to adopt/abandon each type of \textbf{A.I.}, one can compute, for a population of size $n$ with $k$ A.I. adopters, the probability to increase and decrease the number $k$ by 1 at each time-step ($T^{+}(k)$ and $T^{-}(k)$, respectively). These transition probabilities can be used to assess the most probable direction of evolution, given by the so-called imitation gradient \cite{pacheco2008evolutionary,santos2011risk}, $G(k)$, as:
\begin{equation}
        {G(k)} = T^+(k) - T^-(k) 
\end{equation}
where
\begin{eqnarray}
        T^+(k) &=& \frac{n - k}{n}\frac{k}{n}p(f_{H} , f_{A.I.})\tau(f_{H})\\
        T^-(k) &=& \frac{k}{n}\frac{n - k}{n}p(f_{A.I.} , f_{H})
\end{eqnarray}

Being $f_{A.I.}$ the average fitness of the \textbf{A.I.} population and $f_{H}$ the average fitness of the \textbf{H} population. Importantly, we assume that \textbf{H} can adopt an \textbf{A.I.} system only when $f_{H}$ is above a given set Price ($P$), a constraint introduced through $\tau(f_{H})$, given by
\begin{equation}
\tau(f_{H}) = 
\begin{cases} 
      1 &  f_{H} \geq {P} \\
      0 & f_{H} < {P} 
\end{cases}
\end{equation}

When $G(k)>0$ ($G(k)<0$), time evolution is likely to act to increase (decrease) the number of A.I. adopters.
When $G(k)=0$, then we obtain a finite population analogue of a fixed point of a population dynamics in infinite populations \cite{pacheco2008evolutionary,santos2011risk}.

\end{appendix}

%% file: asym.tex
\section{Asymmetric opportunities}
\label{sec:asym}

In the previous sections we show that in some cases, a population of $100\%$ \textbf{A.I.} could improve the overall fitness (Utilitarian  \textbf{A.I.}) when compared with a fully \textbf{H} population, or that the fitness would remain the same with a small reduction in inequality (e.g., Selfish  \textbf{A.I.}). The first case is never reached as an equilibrium, but the second one is achieved when all A.I. types are present and there is no adoption cost (see Fig.~\ref{fig:num_ai_no_cost}). This result might lead us to conclude that, if \textbf{A.I.} is regulated such that everyone is forced to use a particular type of A.I., then both individual and societal fitness would be improved. Moreover, if we force everyone to use the same type of \textbf{A.I.}, individuals will have the same average fitness but with less inequality.

We now perform a complementary analysis to study this fully \textbf{A.I.} populated world and see if other problems may occur. We note that in many cases some people tend to behave in a parochial/discriminatory way \cite{bernhard2006parochial}. This results in behaviors where people prone to parochialism tend to assist people in their own group, while obstructing people from other groups. We do not discuss how such behaviors emerge, for this refer to, e.g., \cite{choi2007coevolution,garcia2011evolution}.

We consider two human populations, $P_1$ and $P_2$, with sizes $s_1$ and $s_2$, and as before we consider that they are playing repeatedly a stochastic game. For this study we consider a repeated prisoner's dilemma game. The ratio of the first population is $b=\frac{s_1}{s_1+s_2}$ and both populations make mistakes, respectively $n_1$ and $n_2$ of the time. The payoff for each individual $i$ is given by $U_i(a_1,a_2)$ where $a_1$ and $a_2$ are the actions chosen by each one.

We further assume that inside each human population each member behaves in the same way, and so we have 4 different combinations of strategies. Either (C)ooperate or (D)efect against their own population and (C) or (D) against the other population. As inside a population everyone behaves in the same way, it is trivial to verify that the best strategy  is to cooperate (C) inside their own population (both individually and for that population) and so we will focus on analyzing the strategies against the other population.  

Without noise we have:
\begin{multline}
U_1(a_1,a_2) =\\ b U_1(C,C) + (1-b) U_1(a_1,a_2)
\end{multline}
and
\begin{multline}
U_2(a_1,a_2) =\\ (1-b) U_2(C,C) + b U_2(a_1,a_2)
\end{multline}
When we consider the mistakes that humans do, the payoff matrix becomes:
\begin{multline}
U_1(a_1,a_2) =\\
b [(1-n_1),\ n_1] U_1(C,C) [(1-n_1),\ n_1]^T\\
+ (1-b) P(a,n) U_1(a_1,a_2) P(a,n)^T
\end{multline}
where $P(a,n) = [\delta_{a=C}(n),\ \delta_{a=D}(n)]$, with $\delta_{a=A}(n)=n$ if $a=A$ and $\delta_{a=A}(n)=1-n$ if $a \neq A$, represents a level $n$ of noise choosing $1-n$ times the correct action and $n$ the wrong one. $U_2$ has a similar structure.

The final value of the difference in payoff between the two populations is:

\begin{multline}
U_1-U_2 = \\
\left[
\begin{array}{cc}
U_1(C,C)-U_2(C,C)&U_1(C,D)-U_2(C,D)\\
U_1(D,C)-U_2(D,C)&U_1(D,D)-U_2(D,D)\\
\end{array}
\right]\\
=\left[
\begin{array}{c}
n_{1} - n_{2}  \\
- 2 b n_{1} + b - n_{1} - n_{2} + 1 
\end{array}\right]\\
\left[
\begin{array}{c}
- 2 b n_{2} + b + n_{1} + 3 n_{2} - 2\\
- 2 b n_{1} - 2 b n_{2} + 2 b - n_{1} + 3 n_{2} - 1
\end{array}
\right]
\end{multline}

For this case we can see that a Nash equilibrium exist for Defect-Defect, assuming the trivial condition that $n_2<.5 \wedge n_1<.5$ (the amount of mistakes is less than $50\%$).

We will now see what happens when an A.I. system is introduced in $100\%$ of the population. In this case both $P_1$ and $P_2$ will adopt the same \textbf{A.I.} type. In Sec.~\ref{sec:HVSR} we discussed several ways to model the differences between humans and A.I. and consider that A.I. systems make better decisions because they have knowledge about the correct payoff matrix. Now we consider another complementary aspect of A.I.: It can allow to reduce the number of errors committed by humans. Similar qualitative results can be obtained with both ways of modelling \textbf{A.I.} and so in this simulation we use the alternative one. If humans behave parochially, an A.I. system being value aligned with its user it will also behave parochially while making less mistakes. This might be a problem as already discussed by \cite{yampolskiy2013artificial}, where an A.I. systems with super-human intelligence might lead to an increase in the number of problems that are not detected by regular human intelligence.

Considering that A.I. systems reduce errors with the rate $c$, then the new error rates become $(1-c)n_1$ and $(1-c)n_2$. The advantage in payoff, how much more payoff is obtained due to the use of A.I., is $c n_1(2b + 1)$ for $P_1$ and $c n_2(3-2b)$ for $P_2$. If $n=n_1=n_2$, the difference between the two is thus $(4b-2)nc$, increasing with $b$, $n$ and $c$.

\begin{figure}
	\centering
		\includegraphics[width=0.45\textwidth]{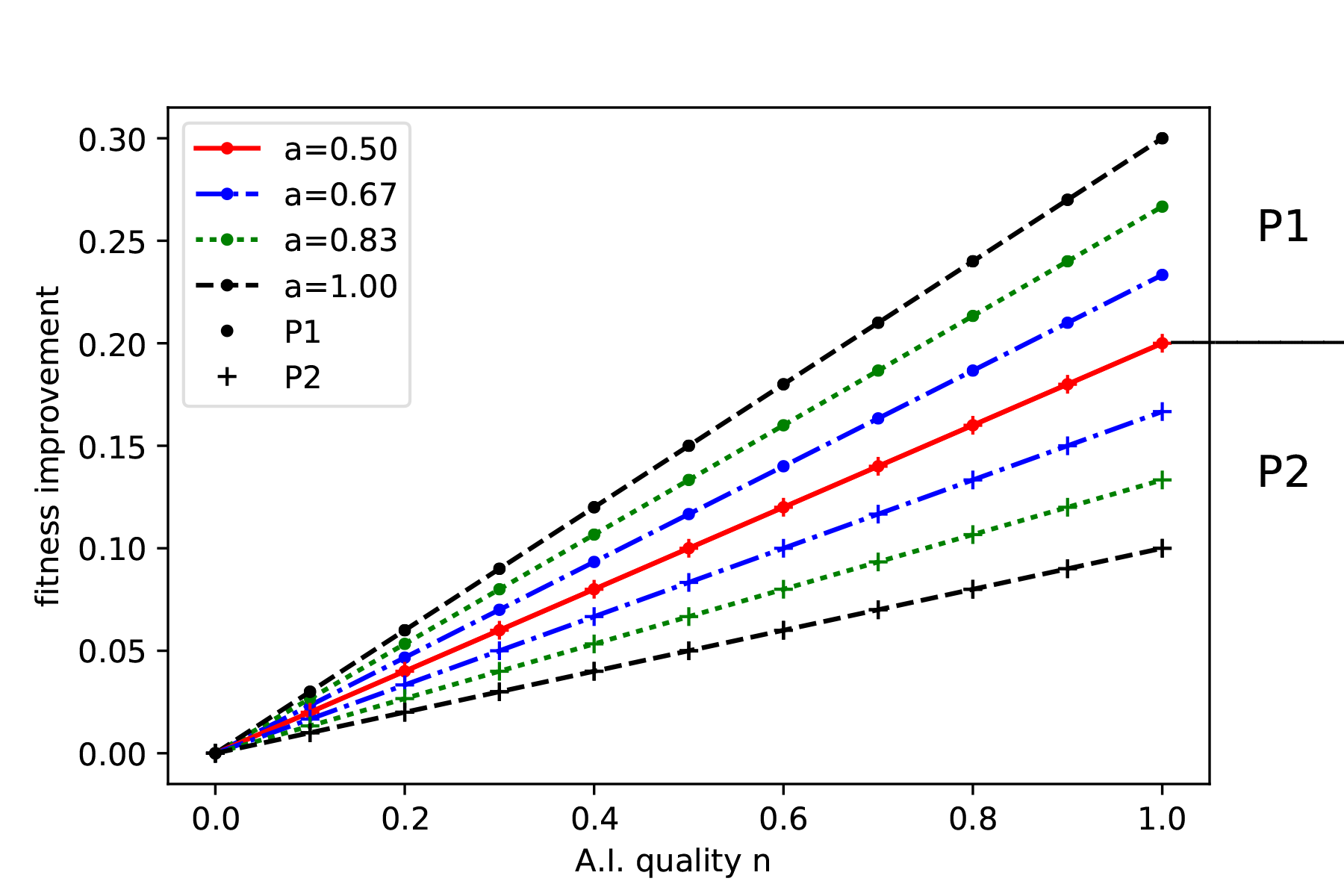}
	\caption{Improvement of fitness for two populations with asymmetric opportunities, when both receive an \textbf{A.I.} system that allows them to commit less errors. Both populations have the same base error rate $n_1=n_2=n$. In the case of equality $b=0.5$ (red line) they both obtain the same gain. When $P_1$ becomes the majority, their gain increases faster with the improvement in quality of the \textbf{A.I.} system (increasing $n$ and with the increase of population majority (increase in $a$). Result shown for $b=\{.5,.67,.83,1.0\}$.}
	\label{fig:prisdilemincAI}
\end{figure}

In Fig.~\ref{fig:prisdilemincAI} we show how much more fitness each population obtains when we increase the quality of the A.I. systems used by both population simultaneously. We see that the minority population always has less benefit from the use of A.I. than the majority population. This effect increases with the increase of the population difference and with the improvement on the quality of the A.I. system (higher $c$).

This result shows that when an entire (parochial) population uses an advanced A.I. system, there will be an increase the inequality between different parts of the population, being this inequality exacerbated by less error prone A.I. systems.

